\documentclass[twocolumn,prb,amssymb,showpacs]{revtex4}
\usepackage{graphicx}

\begin{document}

\title{Hole-hole interaction in a strained In$_x$Ga$_{1-x}$As two dimensional system}

\author{G.~M.~Minkov}
\author{A.~A.~Sherstobitov}
\affiliation{Institute of Metal Physics RAS, 620219 Ekaterinburg,
Russia}

\author{A.~V.~Germanenko}
\author{O.~E.~Rut}
\author{V.~A.~Larionova}
\affiliation{Institute of Physics and Applied Mathematics, Ural
State University, 620083 Ekaterinburg, Russia}

\author{B.~N.~Zvonkov}

\affiliation{Physical-Technical Research Institute, University of
Nizhni Novgorod,  603600 Nizhni Novgorod, Russia}

\date{\today}

\begin{abstract}
The interaction correction to the conductivity of 2D hole gas in
strained GaAs/In$_x$Ga$_{1-x}$As/GaAs quantum well structures was
studied. It is shown that the Zeeman splitting, spin relaxation and
ballistic contribution should be taking into account for reliable
determination of the Fermi-liquid constant  $F_0^\sigma$. The proper
consideration of these effects allows us to describe both th
temperature and magnetic field dependences of the conductivity and
find the value of $F_0^\sigma$.
\end{abstract}

\maketitle
\section{Introduction}
\label{sec1} The transport properties of two dimensional (2D)
systems reveal the intriguing features. One of the feature is a
metallic-like temperature dependence of the resistivity
($\partial\rho/\partial T>0$) at low temperature in some kind of 2D
systems, e.g., in n-Si MOSFET and 2D hole gas in
Al$_x$Ga$_{1-x}$As/GaAs and Ge$_{1-x}$Si$_x$/Ge structures (see
Refs. \onlinecite{Sav04} and \onlinecite{Pud03} and references
therein). As a rule such a behavior is observed in the structures
with large value of the gas parameter $r_s=\sqrt{2}/(a_B k_F)$,
where $a_B$ and $k_F$ are the Bohr radius and Fermi quasimomentum,
respectively, which characterizes the electron-electron ({\em e-e})
or hole-hole ({\em h-h}) interaction strength.

Up to now there is not conventional opinion whether the
metallic-like temperature dependence of the conductivity attests on
quantum phase transition or it results from the interaction
correction to the conductivity. For low $r_s$-values and within the
diffusion regime ($T\tau\ll 1$, where $\tau$ is the transport
relaxation time and $\hbar=k_B=1$), the interaction correction is,
as a rule, negative  and increases in absolute value with the
lowering temperature. However, for the intermediate ($T\tau\simeq
1$) and ballistic ($T\tau\gg 1$) regimes this correction can change
sign leading to the metallic behavior of the resistivity
($\partial\rho/\partial T>0$).\cite{Zala01} Such effect crucially
depends on the value of the Fermi-liquid constant $F_0^\sigma$ (see
Figs.~7 and 8 in Ref.~\onlinecite{Zala01}), therefore the
experimental determination of the interaction correction to the
conductivity and the value of $F_0^\sigma$ is a central point of
numerous papers during the last few years.

\begin{figure}
\includegraphics[width=0.8\linewidth,clip=true]{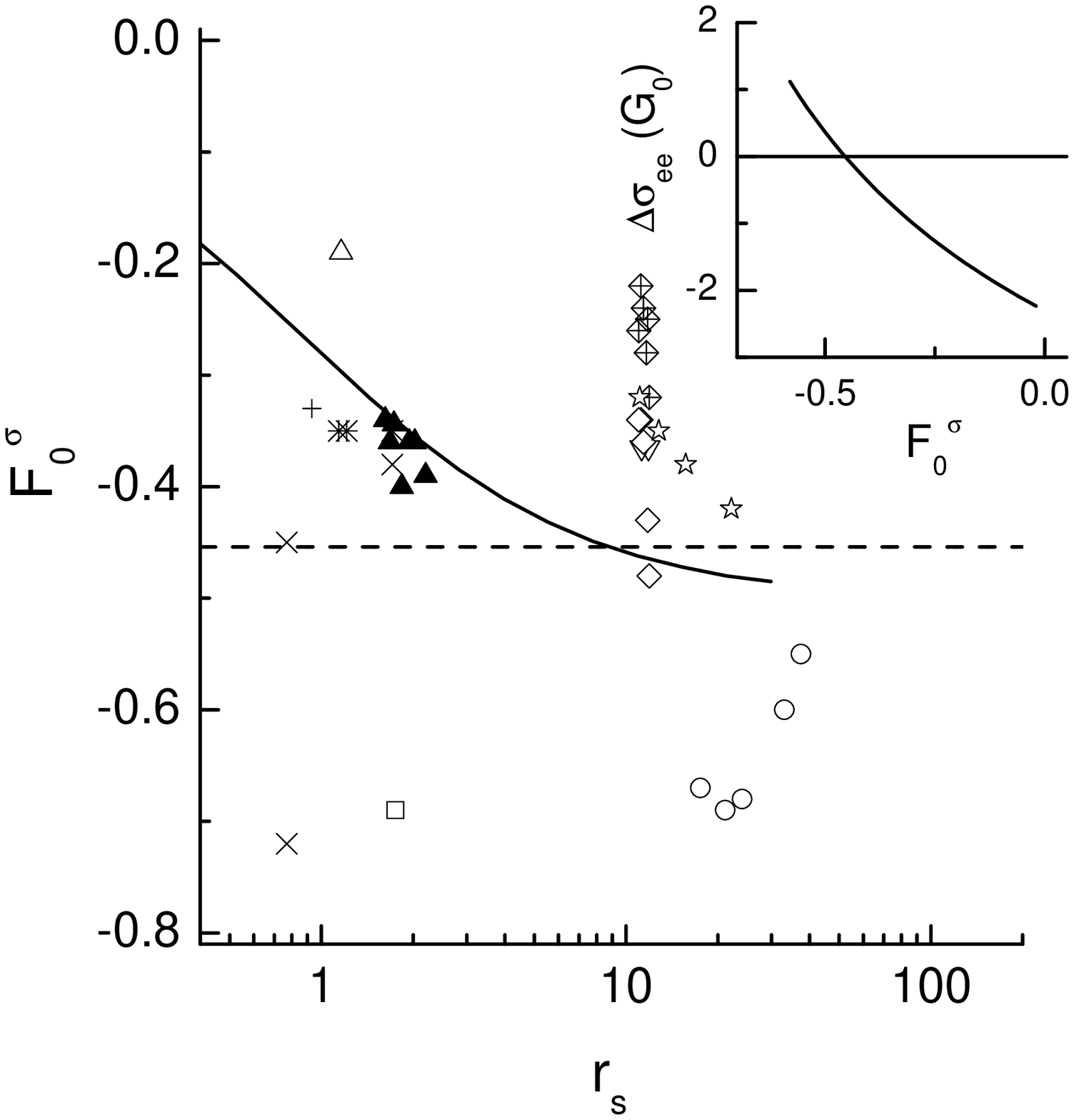}
\caption{The value of $F_0^\sigma$ determined experimentally for
p-type structures Al$_x$Ga$_{1-x}$As and Ge$_{1-x}$Si$_x$/Ge:
($\circ$) Ref.~\onlinecite{Noch03}; ($+$) Ref.~\onlinecite{Emel93};
($\bigtriangledown$) Ref.~\onlinecite{Sim00};
($\triangleleft,\,\triangleright$) Ref.~\onlinecite{Yas04};
($\star$) Ref.~\onlinecite{Li02}; ($\square$)
Ref.~\onlinecite{Arap04}; ($\vartriangle$) Ref.~\onlinecite{Arap99};
($\times$) Ref.~\onlinecite{Col02}; ($\ast$)
Ref.~\onlinecite{Senz00}; ($\blacktriangle$) our data. Solid line is
the theoretical $r_s$-dependence of $F_0^\sigma$,
Ref.~\onlinecite{Zala01}. The dashed line shows $F_0^\sigma=-0.454$
when the correction is equal to zero in the diffusion regime. The
inset shows the diffusion {\em h-h} correction at tenfold changing
of the temperature as a function of $F_0^\sigma$. }
 \label{f0}
 \end{figure}

Unlike  the case of structures with the electron 2D gas there are
some difficulties in extraction  of the interaction correction to
the conductivity in p-type structures with complex valence band,
especially, for the low hole density. First of all, the large value
of perpendicular $g$-factor (which responsible for the Zeeman
splitting in magnetic field perpendicular to the structure plane)
leads to an additional temperature dependence and appearance of
magnetic field dependence of the interaction correction
$\delta\sigma_{xx}^{ee}$ even in the diffusion regime. Besides, for
the high mobility 2D hole gas, the parameter $T\tau$ is usually
greater than unity for all available temperatures, the ballistic
contribution is important and, therefore, the interaction correction
contributes not only to $\sigma_{xx}$, but to $\sigma_{xy}$ as well.
Strong anisotropy of $g$-factor (we mean the strongly different
Zeeman splitting in in-plane and perpendicular magnetic field) makes
it difficult to determine the reliable value of $F_0^\sigma$ from
in-plane magnetoresistance experiments. Most likely just these facts
lead to a very large scatter in the $F_0^\sigma$-values determined
for the 2D hole gas in different papers as illustrated by
Fig.~\ref{f0}. Moreover the the value of $F_0^\sigma$ determined
from the different effects in the same structures, namely, from the
temperature dependence of the conductivity at $B=0$ and from the
temperature dependence of $\rho_{xx}$,  occurs significantly
different.\cite{Yas04}

For the first sight, it may appear that the scatter in $F_0^\sigma$
is not so large. However, it should be pointed out that the
$F_0^\sigma$-sensitivity of the {\em h-h} interaction correction is
very strong. So, a variation of $F_0^\sigma$ within the scatter
range leads  to the strong variation in the value of interaction
correction so that even the sign of the correction becomes different
in the diffusion regime. This illustrates by the inset in
Fig.~\ref{f0}, where the changing of this correction at tenfold
temperature variation as a function of  $F_0^\sigma$ is plotted.

From our point of view for the reliable determination of
$F_0^\sigma$ and enucleation of the role of {\em e-e} ({\em h-h})
interaction in the forming of the metallic-like temperature
dependence of the conductivity it is necessary  to study it
thoroughly starting from the conditions, under which the  theories
are attested, i.e., at low enough $r_s$-value and in the
circumstances, where the ballistic contribution is not dominant.

In the present  paper we  study the {\em h-h} interaction correction
to the conductivity for the low-mobility, high-density
[$p=(3.9-7.2)\times 10^{11}$ cm$^{-2}$] $p$-type quantum well
heterostructures with $r_s \simeq 1.5-2.3$ within the temperature
range from $0.4$ to $4.4$~K, when the parameter $T\tau$ lies within
the interval from $0.02$ to $0.8$ that captures the diffusion
region.

\section{Theoretical background}
\label{sec:TB}

Let us begin  with the diffusion regime, $T\tau \ll 1$. For $B=0$,
the {\em e-e} ({\em h-h}) correction to the conductivity gives the
logarithmic  contribution to the conductivity

\begin{equation}
{\delta \sigma^{ee}(T)\over G_0}=\left[1+3\left(1-
\frac{\ln(1+F_0^\sigma)}{F_0^\sigma}\right)\right]\ln{T\tau},
\label{eq1a}
\end{equation}
where $G_0=e^2/(2\pi^2\hbar)\simeq 1.23\times 10^{-5}$
$\Omega^{-1}$, the first term in square brackets  is the exchange or
the Fock contribution while the second one is the Hartree
contribution (the triplet channel). In most cases the value in the
square brackets is positive and at $T\tau \ll 1$ the interaction
correction is localizing and leads to the logarithmic decrease of
the conductivity with the temperature decrease.

The specific feature of the  interaction correction in this regime
is the fact that in magnetic field  it contributes to $\sigma_{xx}$
and does not to $\sigma_{xy}$:\cite{AA85}
\begin{eqnarray}
\delta \sigma^{ee}_{xx}&=&\delta\sigma^{ee}(T) \label{eq10a}\\
\delta \sigma^{ee}_{xy}&=&0.\label{eq10b}
\end{eqnarray}
As shown in Ref.~\onlinecite{Hough82}, Eqs.~(\ref{eq10a}) and
(\ref{eq10b}) remain to be valid in a classically strong magnetic
field. The absence of the interaction contribution  to $\sigma_{xy}$
leads to the parabolic magnetic field dependence of $\rho_{xx}$ at
$\delta\sigma_{xx}^{ee}(T)\ll \sigma_{xx}$:
\begin{equation}
\rho_{xx}(B,T)  \simeq
\frac{1}{\sigma_0}-\frac{1}{\sigma_0^2}\left(1-\mu^2 B^2\right)
\delta\sigma_{xx}^{ee}(T). \label{eq1}
\end{equation}
The following peculiarity of the interaction induced parabolic
negative magnetoresistance is evident. Despite the fact that the
curvature of the parabola $\rho_{xx}(B)$ is temperature dependent,
there is the point at $B=\mu^{-1}$, in which all the parabolas
relating to different temperatures should cross each other.

Note, we will use the exact relationship between resistivity and
conductivity tensor components
\begin{equation}
 \rho_{xx}=\frac{\sigma_{xx}}{\sigma_{xx}^2+\sigma_{xy}^2}.
 \label{eq6}
\end{equation}
because Eq.~(\ref{eq1}) markedly deviates from Eq.~(\ref{eq6})
already at $\delta\sigma_{xx}^{ee}(\mu B)^2\gtrsim (0.05
-0.1)\sigma_{xx}$.

Up to now we neglected the Zeeman splitting. Taking it into account
results in  the appearance of the magnetic field dependence of
$\delta\sigma_{xx}^{ee}$ at $E_z/T\gtrsim 1$ where $E_z=g\mu_BB$.
This is because that the magnetic field suppresses two components of
the triplet channel while the Fock contribution and one component of
the triplet channel remain unchanged. As a result,  the multiplier
$3$ in Eq.~(\ref{eq1a}) should be replaced by $1$ in high magnetic
field, $E_z/T\gg 1$. The general expressions for the magnetic field
dependence of the interaction correction were obtained in
Ref.~\onlinecite{Fin} for the weak interaction and in
Ref.~\onlinecite{Cast} for the arbitrary one. However, they are too
complicated and it is not convenient  to use them in practice. Much
simpler expression, which well approximates these formulae has been
proposed by I.~V.~Gornyi:\cite{Igor}
\begin{eqnarray}
{\delta \sigma^{ee}_{xx}\over G_0}&=&\ln{T\tau}+\left(1-
\frac{\ln(1+F_0^\sigma)}{F_0^\sigma}\right)
\nonumber \\
 &\times & \left[\ln{T\tau}+2
\ln\left(T\tau\sqrt{1+\left(\frac{E_z}{T}\right)^2}\right)\right]
\label{eq2}.
\end{eqnarray}

It is clear that the sensitivity of $\delta\sigma_{xx}$ to the
magnetic field strength via the Zeeman splitting  results in the
more complicated $B$-dependence of the magnetoresistance as compared
with the parabolic one, Eq.~(\ref{eq1}). Nevertheless, the
magnetoresistance curve retains the parabolic-like shape in
classically strong magnetic field.

Note, namely the fact that the {\em e-e} ({\em h-h}) interaction in
the diffusion regime  contributes only to $\sigma_{xx}$ gives a
possibility to  extract reliably this contribution from the
experimental data.

Let us now consider the more general case when both the diffusion
and ballistic contributions are of importance. In the absence of
magnetic field the interaction correction reads\cite{Zala01}
\begin{eqnarray}
{\delta \sigma^{ee}\over G_0}&=&2\pi
T\tau\left[1-\frac{3}{8}f(T\tau)+\frac{3\widetilde{F}_0^\sigma}{1+\widetilde{F}_0^\sigma}\right.\nonumber \\
&\times&
\left.\left(1-\frac{3}{8}t(T\tau,\widetilde{F}_0^\sigma)\right)\right]
\nonumber \\
 &+& \left[1+3\left(1-\frac{\ln(1+F_0^\sigma)}{F_0^\sigma}\right)\right]\ln{T\tau}
\label{eqS0}.
\end{eqnarray}
In explicit form the functions  $f(T\tau)$ and
$t(T\tau,\widetilde{F}_0^\sigma)$ are given in
Ref.~\onlinecite{Zala01}. The relationship between
$\widetilde{F}_0^\sigma$ and $F_0^\sigma$ can be found from the
simultaneous solution of the equations written out in page 6 of
Ref.~\onlinecite{Zala01}. However, it is much simpler to use the
following approximate formula, which accuracy is better than 2~\%
when $F_0^\sigma=-(0.02\ldots 0.5)$:
\begin{equation}
\widetilde{F}_0^\sigma\simeq F_0^\sigma \left[1.25
\left(F_0^\sigma\right)^{0.69}+0.223\right].\label{eqF0s}
\end{equation}

In the presence of magnetic filed, the situation is more complicated
as compared with the purely diffusion regime, because the {\em e-e}
({\em h-h}) interaction gives a contribution not only to
$\sigma_{xx}$ but to $\sigma_{xy}$ as well. Theory for the ballistic
and intermediate regime, when the strong inequality $T\tau\ll 1$ is
not fulfilled,  is developed for classically low perpendicular
magnetic field in Ref.~\onlinecite{Zala01}. The ballistic regime for
the  high magnetic field was considered in two papers,
Refs.~\onlinecite{Gor04} and \onlinecite{Zala02}. However, all the
analytical results in Ref.~\onlinecite{Gor04} are presented for the
limiting cases: $\omega_c\tau\gg 1$, $E_z/T\gg 1$. Under our
experimental conditions these inequalities are not fulfilled.
Therefore we will use the results obtained in
Ref.~\onlinecite{Zala02} where the influence of Zeeman splitting was
studied  for the in-plane magnetic field orientation. According to
this paper and Ref.~\onlinecite{Zala01} one can write the following
expression for the interaction correction in the presence of
parallel magnetic field
\begin{equation}
\delta\sigma^{ee}_\parallel(B,T)=\delta\sigma^{ee}(0,T)+\Delta\sigma_\parallel(B,T),\label{eq25}
\end{equation}
where the first term $\delta\sigma^{ee}(0,T)$ is given by
Eq.~(\ref{eqS0}) and the second one is\cite{Zala02}
\begin{eqnarray}
\Delta\sigma_\parallel(B,T)&= &2\pi G_0
\left[\frac{2\widetilde{F}_0^\sigma}{1+\widetilde{F}_0^\sigma}T\tau
K_b\left(\frac{E_z}
{2T},\widetilde{F}_0^\sigma\right)\right. \nonumber \\[3mm]
 &+&\left.K_d\left(\frac{E_z}
{2\pi T},F_0^\sigma\right)+m(\ldots)\right].\label{eq3}
\end{eqnarray}
Here, the functions $K_b(x,\widetilde{F}_0^\sigma)$ and
$K_d(x,F_0^\sigma)$  given by Eqs.~(12) and (15) from
Ref.~\onlinecite{Zala02}. The function $m(\ldots)$ describes the
crossover between diffusion and ballistic regimes and only slightly
modifies the sum of the first two terms in Eq.~(\ref{eq3}). The
first term in Eq.~(\ref{eq3}) describes the ballistic contribution
and the second one does the diffusion contribution.

To adapt these results for the analysis of data obtained in the
presence of perpendicular magnetic field we will take into account
the following two well known facts relating to the limiting cases.
In the purely diffusion regime, the {\em e-e} ({\em h-h})
interaction contributes to $\sigma_{xx}$ and does not to
$\sigma_{xy}$ as seen from Eqs.~(\ref{eq10a}) and (\ref{eq10b}). In
the  ballistic regime, the role of the interaction reduces to the
renormalization of the transport relaxation time $\tau$. In this
basis, we suppose that the conductivity components for the arbitrary
temperature and magnetic field values with consideration for  the
Zeeman splitting are
\begin{eqnarray}
  \sigma_{xx}&=&\frac{e^2n}{m}\frac{\tau'}{1+(\omega_c\tau')^2}+\nonumber \\ &+&2\pi
  G_0K_d\left(\frac{E_z}
{2\pi T},F_0^\sigma\right)+\delta\sigma^{ee}(T),  \label{eq4a} \\
  \sigma_{xy}&=&\frac{e^2n}{m}\frac{\omega_c\tau'^2}{1+(\omega_c\tau')^2}, \label{eq4b}
\end{eqnarray}
where $\tau'$ is the momentum relaxation time modified by the
ballistic contribution as
\begin{eqnarray}
 \tau'&=&\tau\left\{1+\frac{T}{E_F}\left[1-\frac{3}{8}f(T\tau)\right.\right.\nonumber\\
& +&\frac{3\widetilde{F}_0^\sigma}{1+\widetilde{F}_0^\sigma}
  \left(1-\frac{3}{8}t(T\tau,\widetilde{F}_0^\sigma)\right)\nonumber \\
  &+&\left.\left.\frac{2\widetilde{F}_0^\sigma}{1+\widetilde{F}_0^\sigma}K_b\left(\frac{E_z}
{2T},\widetilde{F}_0^\sigma\right)\right]\right\}.
 \label{eq5}
\end{eqnarray}
The magnetic field and temperature dependences of $\rho_{xx}$ are
obtained in ordinary way with the help of Eq.~(\ref{eq6}).

We realize that this way is not rigorous. However, to the best of
our knowledge there is not exact solution accounting for all the
peculiarities mentioned above.

To conclude this section, the  interaction correction has to reveal
itself in the different aspects: (i) it should influences the
temperature dependence of the conductivity in the absence of the
magnetic field; (ii) it specifically changes the temperature and
magnetic field dependences of the conductivity tensor component
$\sigma_{xx}$ and $\sigma_{xy}$ in the presence of magnetic field.
It is clear that only in the case when all the mentioned above
dependences are well described in the framework of a common model
one can consider the result as conclusive  and the value of the
parameter $F_0^\sigma$, which determines the value of interaction
correction, can be considered as found reliably.

\section{Experiment}

We have studied the interaction correction in the heterostructures
GaAs/In$_x$Ga$_{1-x}$As/GaAs grown by metal-organic vapor phase
epitaxy on semi-insulator  GaAs substrate. The lattice mismatch
between In$_x$Ga$_{1-x}$As and GaAs results in biaxial compression
of the quantum well. Two types of heterostructures were studied. The
structures of the first type, 3855 and 3857, consist of a
$250$~nm-thick undoped GaAs buffer layer, carbon $\delta$-layer, a
7~nm spacer of undoped GaAs, a 10~nm In$_{0.2}$Ga$_{0.8}$As well, a
7~nm spacer of undoped GaAs, a carbon $\delta$-layer   and 200~nm
cap layer of undoped GaAs. The structure of the second type, 3951,
was analogous, the only difference was the wider spacer, 15~nm, and
as sequence the higher momentum relaxation time.   The samples were
mesa etched into standard Hall bars and then an Al gate electrode
was deposited by thermal evaporation onto the cap layer through a
mask. Varying the gate voltage $V_g$  we were able to change the
hole density and mobility (transport relaxation time) in the quantum
well (see Table \ref{tab1}).

For the quantitative interpretation of the experimental results one
needs to know the effective mass and $g$-factor. The hole effective
mass in the structures investigated has been experimentally
determined from the temperature dependence of the amplitude of the
Shubnikov-de Haas oscillations. It is equal to $(0.160\pm
0.005)$\,$m_0$ and does not depend on the hole density. This value
differs appreciably from the theoretical one, which can be easily
estimated. It can be done with the help of the Luttinger-Kohn
Hamiltonian,\cite{Lut55} which includes the terms responsible for
the strain.\cite{Bir72} Since the Fermi energy, $E_F\simeq
5-10$~meV, in our case  is much less than the strain induced
splitting of the valence band, $2|S|\simeq (80-90)$~meV,\cite{Min05}
the in-plane hole effective mass should be equal to
$m_h=(\gamma_1+\gamma_2)^{-1}m_0$, where $\gamma_i$, $i=1,2$, are
the Luttinger parameters. If one supposes that the Luttinger
parameters of the solid solution In$_{\rm x}$Ga$_{\rm 1- x}$As are
$\gamma_i(x)=\left(x/ \gamma_i^{\rm InAs}+(1-x)/\gamma_i^{\rm
GaAs}\right)^{-1}$, and uses the values of $\gamma_i^{\rm InAs}$ and
$\gamma_i^{\rm GaAs}$ from Ref.~\onlinecite{Vurg01}, we obtain
$m_h/m_0=(\gamma_1+\gamma_2)^{-1}\simeq 0.1$ for
In$_{0.2}$Ga$_{0.8}$As, which is significantly less than the
experimental value. Such a discrepancy was already
reported\cite{Lanc95,Lin95} and, to the best of our knowledge,  has
not an adequate explanation up to now.

As for the $g$-factor, its experimental value is unknown for holes
in strained In$_x$Ga$_{1-x}$As quantum wells. Theoretical value of
$g$-factor for the states with small energy, $\epsilon\ll 2|S|$, is
$g=6\,\kappa$, where $\kappa$ is the Luttinger parameter responsible
for the spin splitting. The use of interpolation formula
$\kappa(x)=\left(x/ \kappa^{\rm InAs}+(1-x)/\kappa^{\rm
GaAs}\right)^{-1}$ gives $g\simeq 8.5$. Remembering the difference
between experimental and theoretical values of $m_h$ we suppose that
the ratio of the cyclotron to Zeeman energies should in reality be
the same as predicted theoretically. Thus, we will use $g=5$ for
In$_{0.2}$Ga$_{0.8}$As.

\begin{table}[b]
\caption{The parameters of structures investigated} \label{tab1}
\begin{ruledtabular}
\begin{tabular}{ccccc}
Structure & $V_g$ (V) &$p$ ($10^{11}$ cm$^{-2}$) & $\tau$ ($10^{-13}$ s)  \\
 \colrule
 3855          &$-0.75$   &$6.3$   &$5.8$\\
 3857          &$1.50$   &$7.2$   &$5.9$ \\
              &$2.00$   &$6.7$   &$5.4$  \\
               &$2.65$   &$5.6$   &$3.7$ \\
 3951          &$0.00$   &$5.0$   &$13$  \\
                &$0.75$   &$4.6$   &$11$ \\
                &$1.25$   &$3.9$   &$9$  \\
 \end{tabular}
\end{ruledtabular}
\end{table}

\begin{figure}
\includegraphics[width=\linewidth,clip=true]{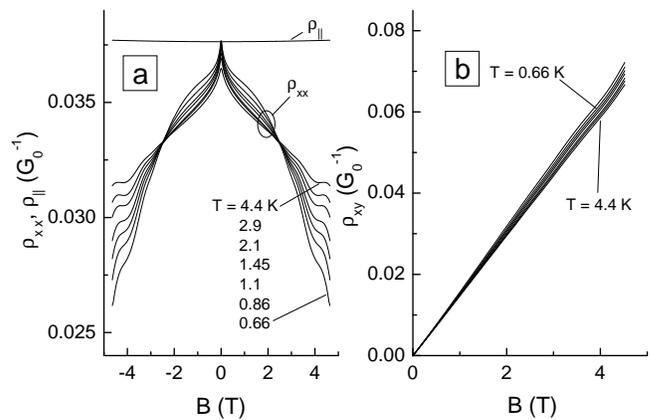}
\caption{The magnetic field dependences of $\rho_{xx}$,
$\rho_\parallel$ (a), and $\rho_{xy}$ (b) measured at different
temperatures for structure 3857 at $V_g=2.65$ V. } \label{f1}
\end{figure}

Now we are in position to consider the experimental data. The role
of {\em h-h} interaction in the structures investigated is evident
from the magnetic field dependences of  $\rho_{xx}$, presented for
one of the structures in Fig.~\ref{f1}. It is clearly seen that
following the sharp magnetoresistance in low magnetic field,
$B<(0.1-0.5)$~T, which results from suppression of the interference
quantum correction,\cite{Min05} the parabolic-like negative
magnetoresistance  is observed. Such a behavior, as discussed in
previous section, shows that the diffusion part of the interaction
correction is dominant.\cite{footn}

\begin{figure}
\includegraphics[width=0.8\linewidth,clip=true]{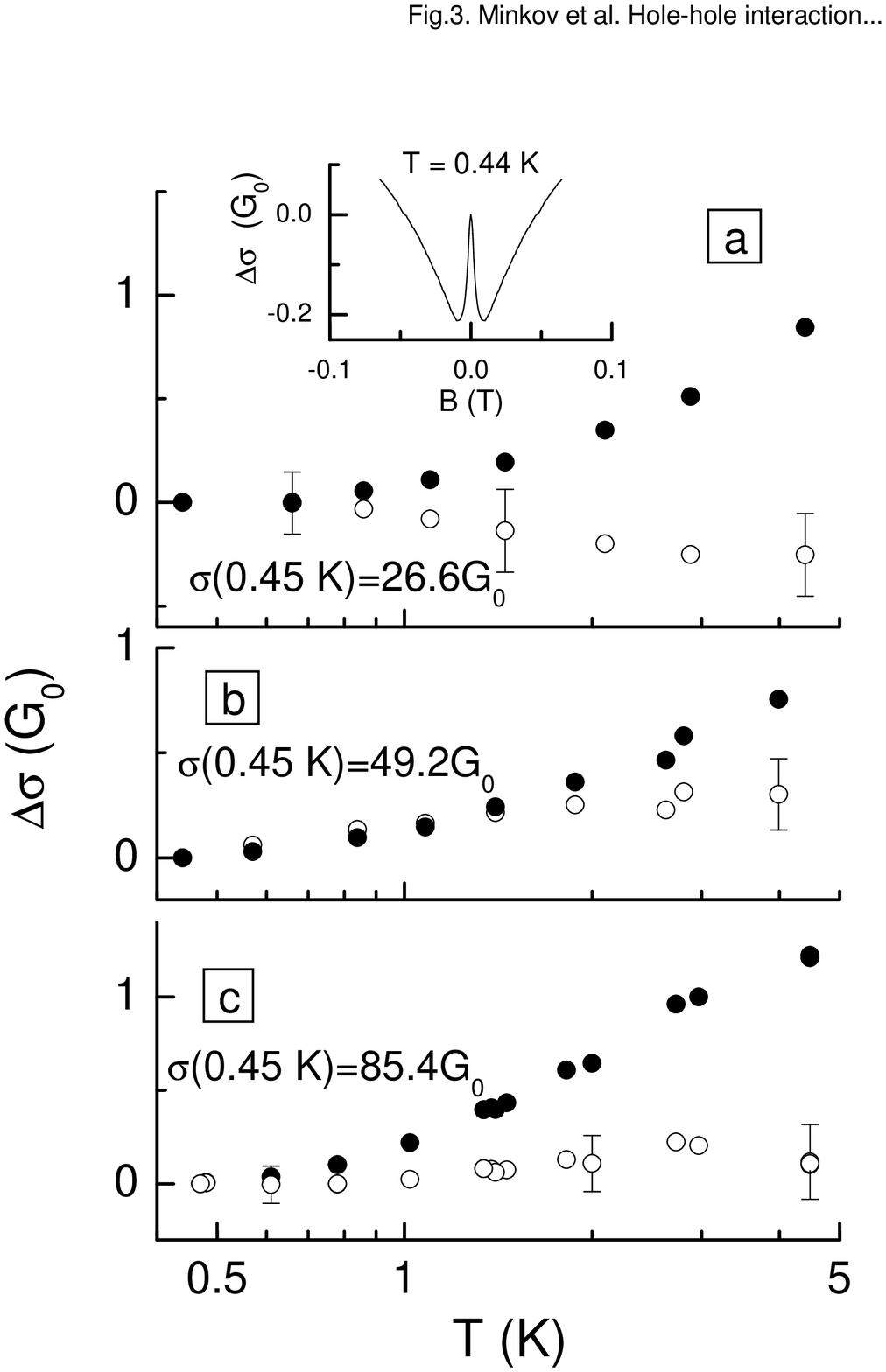}
\caption{The temperature dependences of
$\Delta\sigma=\sigma(T)-\sigma(0.45\text{ K})$ for structure 3857 at
$V_g=2.65$ V (a) and $V_g=2$ V (b), and 3951 at $V_g=0$ (c). The
solid symbols are the original experimental data. The open symbols
are the same data after subtraction of the interference correction
(see text). The inset in (a) shows the low-field magnetoresistance.}
 \label{f2}
\end{figure}

Let us firstly  analyze the experimental data for $B=0$. The
temperature dependence of the conductivity for two heterostructures
with different electron density is presented in Fig.~\ref{f2}. A
strong deviation of the temperature dependence of $\sigma$  from the
logarithmic law is a result of  the spin relaxation, which affects
the logarithmic behavior of the interference correction. An
importance of the spin relaxation is evident from the low-field
magnetoconductance in which it reveals itself as characteristic
antilocalization minimum (see inset in Fig.~\ref{f2}a). The results
of detailed studies of this correction for the structures
investigated were published in Ref.~\onlinecite{Min05}. It has been
shown that the Hikami-Larkin-Nagaoka expression\cite{Hik80} well
describes  the interference correction that allows us to find the
gate-voltage and temperature dependences of the phase and spin
relaxation times and subtract the interference correction from the
experimental $\sigma$-vs-$T$ dependence. The results are presented
in Fig.~\ref{f2} by open symbols. Surprisingly, after such
subtraction the conductivity becomes temperature independent
practically. Thus, we observe the discrepancy: the {\em h-h}
interaction correction to the conductivity reveals itself in the
magnetoresistance curves (see Fig.~\ref{f1}), but does not in the
temperature dependence of the conductivity at $B=0$ (see
Fig.~\ref{f2}).
One of the reasons of this discrepancy can be the ballistic
contribution, which  radically affects the temperature dependence of
the conductivity at $B=0$ for some $F_0^\sigma$ values as mentioned
in Section \ref{sec1}.

\begin{figure}
\includegraphics[width=0.8\linewidth,clip=true]{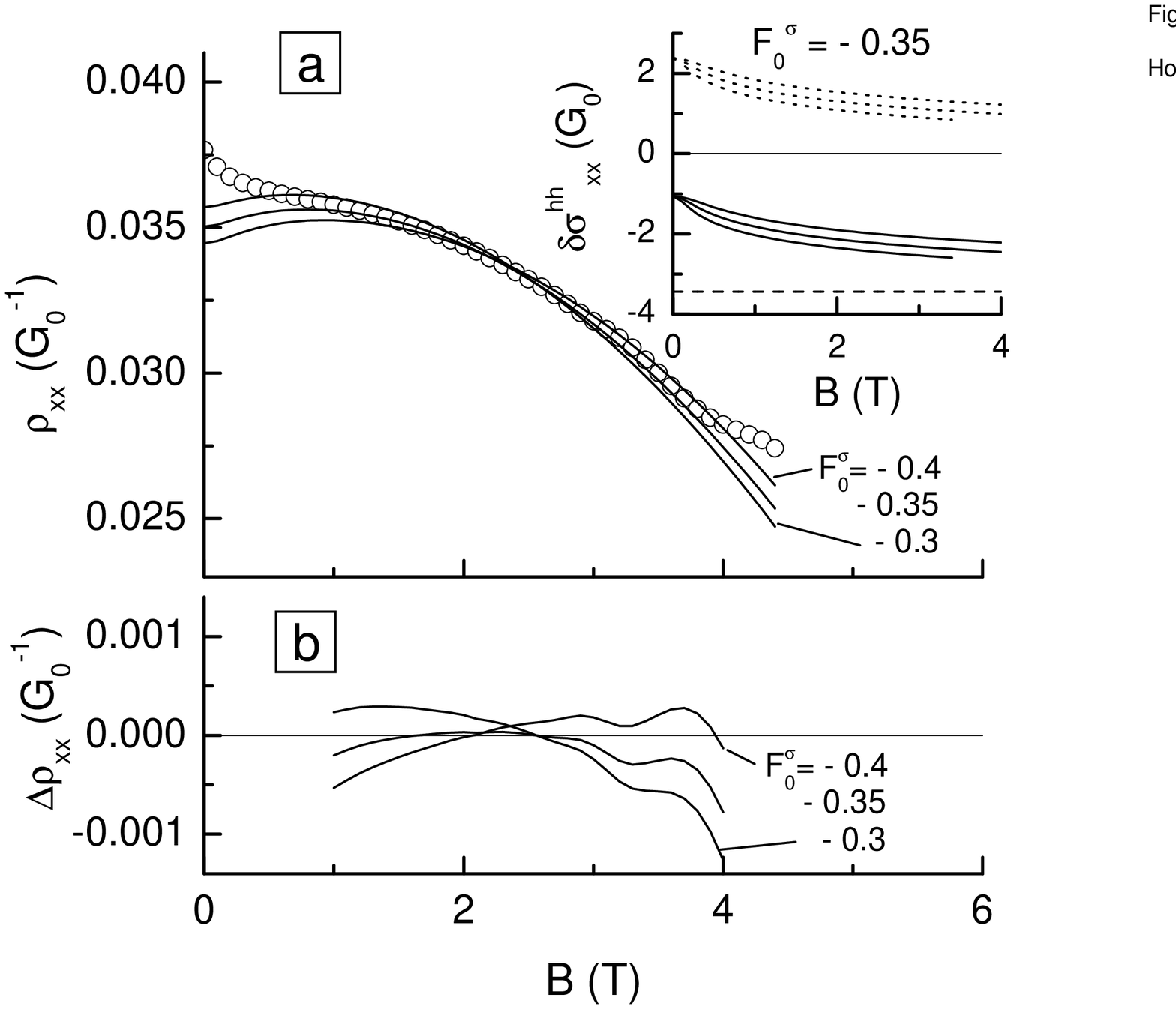}
\caption{The magnetic field dependences of $\rho_{xx}$ (a) and
differences between experimental and calculated $\rho_{xx}$-vs-$B$
curves (b) for structure 3857, $V_g=2.65$ V, $T=0.66$~K. The symbols
are the experimental data, the curves are theoretical dependences
calculated with different $F_0^\sigma$-values. The inset shows the
Fock contribution (dashed line), the Hartree contribution (dotted
curves), and the total correction (solid curves) calculated with
different values of $g$-factor and plotted as a function of magnetic
field: $g=3$, $5$, and $8$ for curves from top to bottom.}
 \label{f3}
\end{figure}

To estimate the value of $F_0^\sigma$ let us analyze the
magnetoresistance at lowest temperature where the ballistic
contribution is rather small. In Fig.~\ref{f3}(a)   we  present the
experimental $\rho_{xx}$-vs-$B$ dependences at lowest temperature
together with the calculated ones with the different values of
$F_0^\sigma$. To make the result of the comparison more clear we
present in Fig.~\ref{f3}(b) the differences between the experimental
and calculated curves. The range of the low magnetic field
$B<10B_{tr}\simeq 1.3$~T, where the interference correction is
significant is cut off. On can see that the best agreement occurs at
$F_0^\sigma=-0.35$. Note, the calculated $\rho_{xx}$-vs-$B$ curve
does not depend strongly on the specific value of $F_0^\sigma$.
Besides, the strong dependence on $g$-factor value is absent as
well. The low sensitivity to $F_0^\sigma$ and $g$-factor is a
sequence of the fact that the main contribution to
$\delta\sigma_{xx}^{ee}$ within actual range of the magnetic field
comes from the Fock term [which is independent of $F_0^\sigma$ and
$g$-factor as evident from Eq.~(\ref{eq2})], because the Hartree
contribution is strongly suppressed at $B>2$~T. The written is
illustrated by the inset in Fig.~\ref{f3}.

Thus taking into account the possible uncertainty in the value of
$g$-factor we estimate the accuracy in the determination of
$F_0^\sigma$ as $\pm0.05$. The close  value of $F_0^\sigma$ was
obtained by this way for other gate voltages and other structures
under the condition $T\tau\lesssim 0.05$.

It may appear from Eq.~(\ref{eq3})  that one can determine the
values of $g$-factor and $F_0^\sigma$ from the analysis of
magnetoresistance measured in  in-plane magnetic field. However such
method could be useful being applied to  a system with isotropic
$g$-factor. For our case of 2D hole gas in strained structures, the
$g$-factor is extremely anisotropic. In particular, the in-plane
$g$-factor should be equal to zero for the  states with the energy
significantly less than the strain induced splitting. Really,  the
in-plane magnetoresistance $\rho_\parallel(B)$ is practically absent
as Fig.~\ref{f1}(a) illustrates.

\begin{figure}
\includegraphics[width=0.8\linewidth,clip=true]{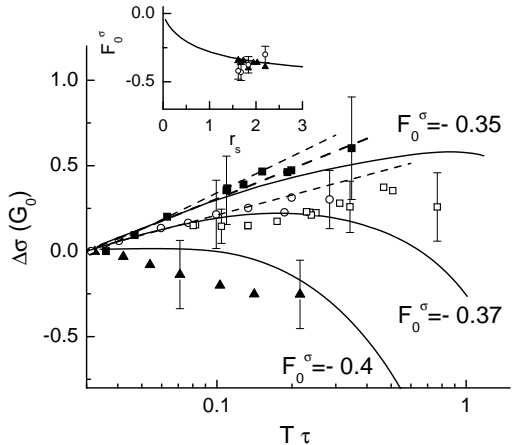}
\caption{The $T\tau$ dependence of the interaction correction. Solid
lines are the  dependences of
$\Delta\sigma(T)=\delta\sigma_{ee}(T)-\delta\sigma_{ee}(0.5\,{\rm
K})$  calculated from Eq.~(\ref{eqS0}) with the different values of
$F_0^\sigma$. The  dashed lines are the diffusion contribution with
the same values of the parameter $F_0^\sigma$.  The symbols are the
experimental data after the substraction of the interference
correction for the structure 3857,  $V_g=2.65$~V ($\blacktriangle$),
$2$~V ($\circ$), and $1.5$~V ($\blacksquare$), and for the structure
3951, $V_g=0$~V ($\square $). The inset shows the values of
$F_0^\sigma$ found from the analysis of the dependences $\sigma(T)$
($\blacktriangle$) and $\rho_{xx}(B)$ ($\circ$) and plotted against
the gas parameter $r_s$. Solid line is the result of
Ref.~\onlinecite{Zala01}.}
 \label{f4}
\end{figure}

Let us now elucidate the role of the ballistic contribution at $B=0$
for $F_0^\sigma=-(0.3\ldots 0.4)$ within actual $T\tau$-range. To do
this we have calculated from Eq.~(\ref{eqS0}) the temperature
dependences of the difference
$\Delta\sigma(T)=\delta\sigma_{ee}(T)-\delta\sigma_{ee}(0.5\,{\rm
K})$ (Fig.~\ref{f4}).One can see that the behavior of
$\Delta\sigma(T)$ strongly depends on value of $F_0^\sigma$ for
actual range of $F_0^\sigma$ and $T\tau$. In addition, the slope of
the  curves calculated with taking into account ballistic
contribution and without that are strongly different even at
$T\tau<0.1$. This  means that the ballistic contribution is
important for actual values of $F_0^\sigma$ and $T\tau$, in
particular the total interaction correction becomes temperature
independent up to $T\tau=0.4$  at $F_0^\sigma=-0.37$. In the same
figure we present the experimental data obtained for the different
gate voltages and different structures. Comparing them with the
theoretical curves we have found the values of $F_0^\sigma$ and
plotted them  in the inset in Fig.~\ref{f4} as a function of $r_s$.
One can see that the experimental data lie close to the theoretical
curve.\cite{Zala01} In the same inset we show the values of
$F_0^\sigma$ determined from the magnetoresistance treatment. Some
difference between $F_0^\sigma$-values obtained by the different
ways can be sequence of the classical magnetoresistance
mechanisms,\cite{Bask78,Bob95,Dmit01,Mir01} which are not essential
in our case, but, nevertheless,  can influence the shape of the
$\rho_{xx}$-curve.

Thus, we have resolved the apparent contradiction between the
clear manifestation of the interaction correction  in the form of
the parabolic-like negative magnetoresistance and the absence of
interaction contribution in the temperature dependence of the
conductivity at $B=0$.  It has been shown that such a behavior
results from the  competition between the diffusion and ballistic
contributions, which have the opposite signs of $d\sigma/dT$ for
actual values of $F_0^\sigma$.

\begin{figure}
\includegraphics[width=\linewidth,clip=true]{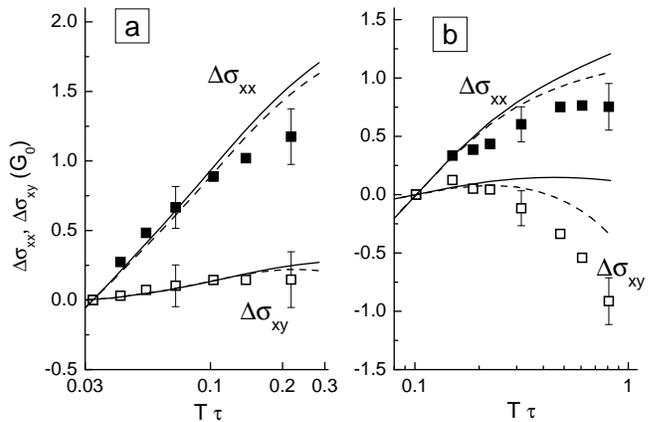}
\caption{The $T\tau$-dependences of
$\Delta\sigma_{xx}=\sigma_{xx}(T)-\sigma_{xx}(0.45\text{ K})$ and
$\Delta\sigma_{xy}=\sigma_{xy}(T)-\sigma_{xy}(0.45\text{ K})$ for
structures 3857 at $V_g=2.65$~V (a) and 3855 at $V_g=-0.75$~V (b),
$B=1/\mu$. The curves are calculated from Eqs.~(\ref{eq4a}) and
(\ref{eq4b}) with $F_0^\sigma=-0.35$ (solid lines) and $-0.4$
(dashed lines).}
 \label{f5}
\end{figure}

Above we analyzed the  magnetoresistance in the diffusion regime and
the temperature dependence of the conductivity at $B=0$ where the
theory gave the rigorous predictions. Now let us consider the
temperature dependences of the conductivity tensor components in a
magnetic field in framework of the model considered in Section
\ref{sec:TB}, which reduces the ballistics  to the mean free time
renormalization. The $T\tau$-dependences of
$\Delta\sigma_{xx}=\sigma_{xx}(T)-\sigma_{xx}(0.45\text{ K})$ and
$\Delta\sigma_{xy}=\sigma_{xy}(T)-\sigma_{xy}(0.45\text{ K})$ at
$B=1/\mu$ for two structures  are presented in Fig.~\ref{f5}. This
value of the magnetic field is chosen specially. It easy to show
that the component $\sigma_{xx}$ at $B=1/\mu$ depends on the
diffusion contribution only, while $\sigma_{xy}$ depends on the
ballistic one. In the same figure the theoretical curves calculated
from Eqs.~(\ref{eq4a}) and (\ref{eq4b}) with $F_0^\sigma=-0.35$ and
$-0.4$ are plotted also. Again, one can see that the experimental
points for both $\sigma_{xx}$ and $\sigma_{xy}$ lie close to the
calculated curves.  Note, the data for the structure 3855 with the
larger values of $T\tau$ deviate from the theoretical curves
somewhat stronger. Probably, it results from not rigorous treatment
of the ballistic contribution in the presence of magnetic field. The
analysis of the temperature dependences of $\sigma_{xx}$ and
$\sigma_{xy}$ for the other magnetic field strength gives
satisfactory agreement with the model also.

We believe that a more detailed analysis of the data in the magnetic
field when the ballistics is important  oversteps the accuracy of
the approximations made above. Besides, one has to perceive that 2D
hole gas is not suitable object for the experimental study of the
interaction correction in the ballistic regime. The reason is the
large value of the Zeeman splitting and high spin relaxation rate,
which strongly modify both the temperature and magnetic field
dependences of the conductivity. From our point of view,  the
interaction correction in intermediate regime, where there are not
firm theoretical predictions, should be studied firstly for the
simpler situation which is realized in 2D electron gas. Such studies
are in progress.

To demonstrate how our values of $F_0^\sigma$ relate to the results
of the previous papers we plot them in Fig.~\ref{f0}. One can see
that the scatter of our experimental points is not larger and they
lie closely to theoretical curve.  In our opinion the large scatter
of the earlier data is  a sequence of the improper consideration of
the Zeeman splitting, the spin relaxation and the ballistic
contribution, which play an important role in 2D hole gas.

\section{Conclusion}

We have studied the interaction correction to the conductivity of 2D
hole gas in strained GaAs/In$_x$Ga$_{1-x}$As/GaAs quantum well
structures. We have shown that for the reliable determination of the
value of the Fermi-liquid constant  $F_0^\sigma$ from the magnetic
field dependences of conductivity in the diffusion regime one has to
take into account the Zeeman splitting, while for its determination
from the temperature dependence of the conductivity in zero magnetic
field the spin relaxation has to be properly accounted. We have
found that for $r_s=1.5-2.3$ the value of $F_0^\sigma$  is equal to
$ -0.35\pm 0.05$.  It has been  shown that for this
$F_0^\sigma$-value the ballistic contribution is important starting
from $T\tau\approx 0.05$.

\subsection*{Acknowledgments}
We thank Igor Gornyi for useful discussion. This work was supported in
part by the RFBR through Grants No. 00-02-16215, No. 01-02-16441 and
No. 01-02-17003, the INTAS through Grant No. 1B290, the Program {\it
University of Russia} through Grants No.~UR.06.01.002, the CRDF through
Grant No. REC-005.


\begin{thebibliography}{99}
\bibitem{footn} The magnetoresistance decreases  in magnitude with
increasing temperature and practically disappears at $T\simeq 30$~K
(not shown in Fig.~\ref{f1}). It points to the fact that the
classical mechanisms,\cite{Bask78,Bob95,Dmit01,Mir01} which should
lead to the temperature independent negative magnetoresistance do
not reveal themselves in our case.
\end{thebibliography}
\end{document}